\documentstyle[aps]{revtex}

\begin{document}
\title{Electrostatic potential of a homogeneously charged square and cube
in two and three dimensions}
\author{Gerhard Hummer}
\address{
Theoretical Biology and Biophysics Group (T-10) and Center for
Nonlinear Studies (CNLS),\\
MS~K710, Los Alamos National Laboratory, Los
Alamos, New Mexico 87545, U.S.A.\\
E-mail: {\tt hummer@t10.lanl.gov}\\
Phone: 1-505-665-1923; FAX: 1-505-665-3493}
\maketitle
\begin{center}
{\em Journal of Electrostatics} (in press, 1995)
\end{center}
\begin{abstract}
A closed form of the electrostatic potential of a homogeneously
charged cube is derived by integration. The exact result is compared
with multipole expansions for the exterior and interior of the cube.
The electrostatic potential of a homogeneously charged square in
two-dimensional electrostatics is also determined.
\end{abstract}

\section{Introduction}\indent

The electrostatic potential of a homogeneously charged cube appears in
theoretical studies of Wigner lattices \cite{Nijboer:88}. In computer
simulations of ionic systems using minimum-image electrostatics, it
determines the electrostatic self-interaction of ions
\cite{Sloth:90,Sorensen:91,Hummer:93,Hummer:95:c,Figueirido:95}. In
Ref.~\cite{Hummer:95:c}, Hummer {\em et al.} presented a simple
calculation of the electrostatic potential at the center of a
homogeneously charged cube. In this work, a closed form of the
electrostatic potential will be determined for arbitrary positions.
This analytic form can be used for the evaluation of lattice sums. It
can also be applied as a correction when electrostatic potentials are
calculated on a grid, assuming that the grid volumes are uniformly
charged rather than carrying a point charge at the center. The
analytic form of the potential will be compared with multipole
expansions \cite{Nijboer:88,Durand:64:cube}

\section{Calculation of the electrostatic potential of a cube}\indent

The electrostatic potential $\phi_c$ of a cube $[-1/2,1/2]^3$ with
charge density one will be calculated by integration. The potential at
a point with Cartesian coordinates $(u,v,w)$ can be written as
\begin{eqnarray}
\phi_c(u,v,w) & = & \int_{-1/2}^{1/2} dx \int_{-1/2}^{1/2} dy
\int_{-1/2}^{1/2} dz \left[ (x-u)^2 + (y-v)^2 + (z-w)^2 \right]^{-1/2}~,
\label{eq:int}
\end{eqnarray}
where Gaussian units are used. $\phi_c$ can be rewritten as
\begin{eqnarray}
\int_{-1/2-u}^{1/2-u} dx \int_{-1/2-v}^{1/2-v} dy \int_{-1/2-w}^{1/2-w} dz
\left( x^2 + y^2 + z^2 \right)^{-1/2}~.
\end{eqnarray}
Summation of the results of partial integration with respect to $x$,
$y$, and $z$ yields a reduction to three two-dimensional integrals,
\begin{eqnarray}
\lefteqn{
2 \phi_c(u,v,w) = \int_{-1/2-u}^{1/2-u} dx \int_{-1/2-v}^{1/2-v} dy
\left[ z \left( x^2 + y^2 + z^2 \right)^{-1/2} \right]^{z=1/2-w}_{z=-1/2-w}
}\nonumber\\
&&\mbox{+ cyclic permutations }
(x,u;y,v;z,w)\rightarrow (y,v;z,w;x,u) \mbox{ and } (z,w;x,u;y,v)~.
\end{eqnarray}
The two-dimensional integrals can be further reduced using
\begin{eqnarray}
\int_{x_0}^{x_1} dx \int_{y_0}^{y_1} dy
\left( x^2 + y^2 + z^2 \right)^{-1/2} & = &
\int_{x_0}^{x_1} dx \; \frac{1}{2} \left[
\ln \frac{(x^2+y^2+z^2)^{1/2}+y}{(x^2+y^2+z^2)^{1/2}-y} \right]
_{y_0}^{y_1}~,
\end{eqnarray}
where $x_0$, $x_1$, $y_0$, and $y_1$ are arbitrary integral boundaries.
The remaining one-dimensional integrals can be calculated using partial
integration and conventional substitution for algebraic integrands,
\begin{eqnarray}
\lefteqn{
\int dx \ln[(x^2+a^2)^{1/2}+b] =
x \ln[ (x^2+a^2)^{1/2} + b ] -x}\nonumber\\ &&+ 2 | a^2 - b^2 |^{1/2}
\;A\!\left[ \frac{x+(x^2+a^2)^{1/2}+b}{|a^2-b^2|^{1/2}}\right] + b \ln [ x
+ ( x^2 + a^2 )^{1/2} ]~,
\end{eqnarray}
where
\begin{eqnarray}
A(x) & = & \left\{ \begin{array}{lll}
                     \arctan(x)       & \mbox{for} & a^2 > b^2\\
                     \mbox{artanh}(x) & \mbox{for} & a^2 < b^2~.
	           \end{array} \right.
\end{eqnarray}
Combining the previous results yields a closed form for the
electrostatic potential of a unit cube:
\begin{eqnarray}
\lefteqn{\phi_c(u,v,w) = \frac{1}{2} \left\{ \frac{1}{2} \sum_{i=0}^{1}
\sum_{j=0}^{1} \sum_{l=0}^{2} (-1)^{i+j} \; c_{i,l} \; c_{j,l+1} \right.}
\nonumber\\
&&\times\ln\frac{
\left[\left( c_{i,l}^2 + c_{j,l+1}^2 + c_{1,l+2}^2 \right)^{1/2} +
c_{1,l+2} \right]^3
\left[\left( c_{i,l}^2 + c_{j,l+1}^2 + c_{0,l+2}^2 \right)^{1/2} -
c_{0,l+2} \right]}{
\left[\left( c_{i,l}^2 + c_{j,l+1}^2 + c_{1,l+2}^2 \right)^{1/2} -
c_{1,l+2} \right]
\left[\left( c_{i,l}^2 + c_{j,l+1}^2 + c_{0,l+2}^2 \right)^{1/2} +
c_{0,l+2} \right]^3} \label{eq:c3}\\
&&\left. +\sum_{i=0}^{1} \sum_{j=0}^{1} \sum_{k=0}^{1} \sum_{l=0}^{2}
(-1)^{i+j+k+1} \; c_{i,l}^2 \; \arctan\frac{
c_{i,l} \; c_{k,l+2}}{
c_{i,l}^2 + c_{j,l+2}^2 + c_{j,l+1}
\left( c_{i,l}^2 + c_{j,l+1}^2 + c_{k,l+2}^2 \right)^{1/2} } \right\}~.
\nonumber
\end{eqnarray}
The integration boundaries are defined as $c_{0,0} = - {1/2} - u$,
$c_{1,0} = {1/2} - u$, $c_{0,1} = - {1/2} - v$, $c_{1,1} = {1/2} - v$,
$c_{0,2} = - {1/2} - w$, and $c_{1,2} = {1/2} - w$. The values of
$l+1$ and $l+2$ in Eq.~(\ref{eq:c3}) are defined modulo 3, {\em i.e.},
$c_{0,3} \equiv c_{0,0}$ etc.\ \ The $\arctan$ function to be used in
Eq.~(\ref{eq:c3}) takes into account the sign of numerator and
denominator and yields results between $-\pi$ and $\pi$ (``atan2'' in
FORTRAN and C).

An immediate consequence of Eq.~(\ref{eq:c3}) is the electrostatic
potential at the center of a unit cube
\begin{eqnarray}
\phi_c(0,0,0) & = & 3 \ln \left( 3^{1/2} + 2 \right) - \frac{\pi}{2}~.
\label{eq:phinull}
\end{eqnarray}
Previous calculations of $\phi_c(0,0,0)$ involved rather elaborate
manipulations \cite{Nijboer:88,Sorensen:91}.

\section{Calculation of the electrostatic potential of a square}\indent

In two-dimensional electrostatics, the charge interaction (Green's
function of the Laplacian) is given by $-\ln r$, where $r$ is the
distance. The electrostatic potential $\phi_s$ of a square
$[-1/2,1/2]^2$ with unit charge density will again be calculated by
integration. $\phi_s$ is also the electrostatic potential of a square
cylinder that is infinitely extended in $z$ direction. The potential
at a point with Cartesian coordinates $(u,v)$ is written as
\begin{eqnarray}
\phi_s(u,v) & = & - \frac{1}{2} \int_{-1/2}^{1/2} dx \int_{-1/2}^{1/2} dy
\ln \left[ (x-u)^2 + (y-v)^2 \right]~.
\end{eqnarray}
Elementary integration yields
\begin{eqnarray}
\phi_s(u,v) & = & - \frac{1}{2} \sum_{i=0}^{1} \sum_{j=0}^{1} \left[
x_i \; y_j \; \ln \left( x_i^2 + y_j^2 \right) - 3 \; x_i \; y_j +
y_j^2 \; \arctan\frac{x_i}{y_j} +
x_i^2 \; \arctan\frac{y_j}{x_i} \right]~, \nonumber\\
\label{eq:c2}
\end{eqnarray}
where $x_0 = -1/2 - u$, $x_1 = 1/2 - u$, $y_0 = -1/2 - v$, and $y_1 =
1/2 - v$. The appropriate $\arctan$ function to be used in
Eq.~(\ref{eq:c2}), yields values between $-\pi/2$ and $\pi/2$
(``atan'' in FORTRAN and C).

\section{Multipole expansion}\indent

The electrostatic potential of a cube can be expanded in ``kubic''
harmonics, {\em i.e.}, harmonic functions with cubic symmetry
\cite{Nijboer:88,Hummer:93,vdLage:47,Slattery:80,Adams:87}. For the
exterior, one obtains
\begin{eqnarray}
\phi_c(\mbox{\bf r}) & = &
\frac { 1 } { r } + C_4 \; K_4(\mbox{\bf r}) \; r^{-9}
+ C_6 \; K_6(\mbox{\bf r}) \; r^{-13} + \cdots ~, \label{eq:mout}
\end{eqnarray}
where $\mbox{\bf r}=(u,v,w)$, $r=|\mbox{\bf r}|$. With
$T_n=u^n+v^n+w^n$, the kubic harmonics of order 4 and 6 can be written
as \cite{Adams:87}
\begin{eqnarray}
K_4(\mbox{\bf r}) & = & T_4 - \frac{3}{5} \; r^4 \\
K_6(\mbox{\bf r}) & = & T_6 - \frac{15}{11} \; T_4 \; r^2 + \frac{30}{77}
\; r^6~.
\end{eqnarray}
For this form, the expansion coefficients are $C_4=-7/192$ and
$C_6=11/192$ \cite{Nijboer:88,Durand:64:cube}.

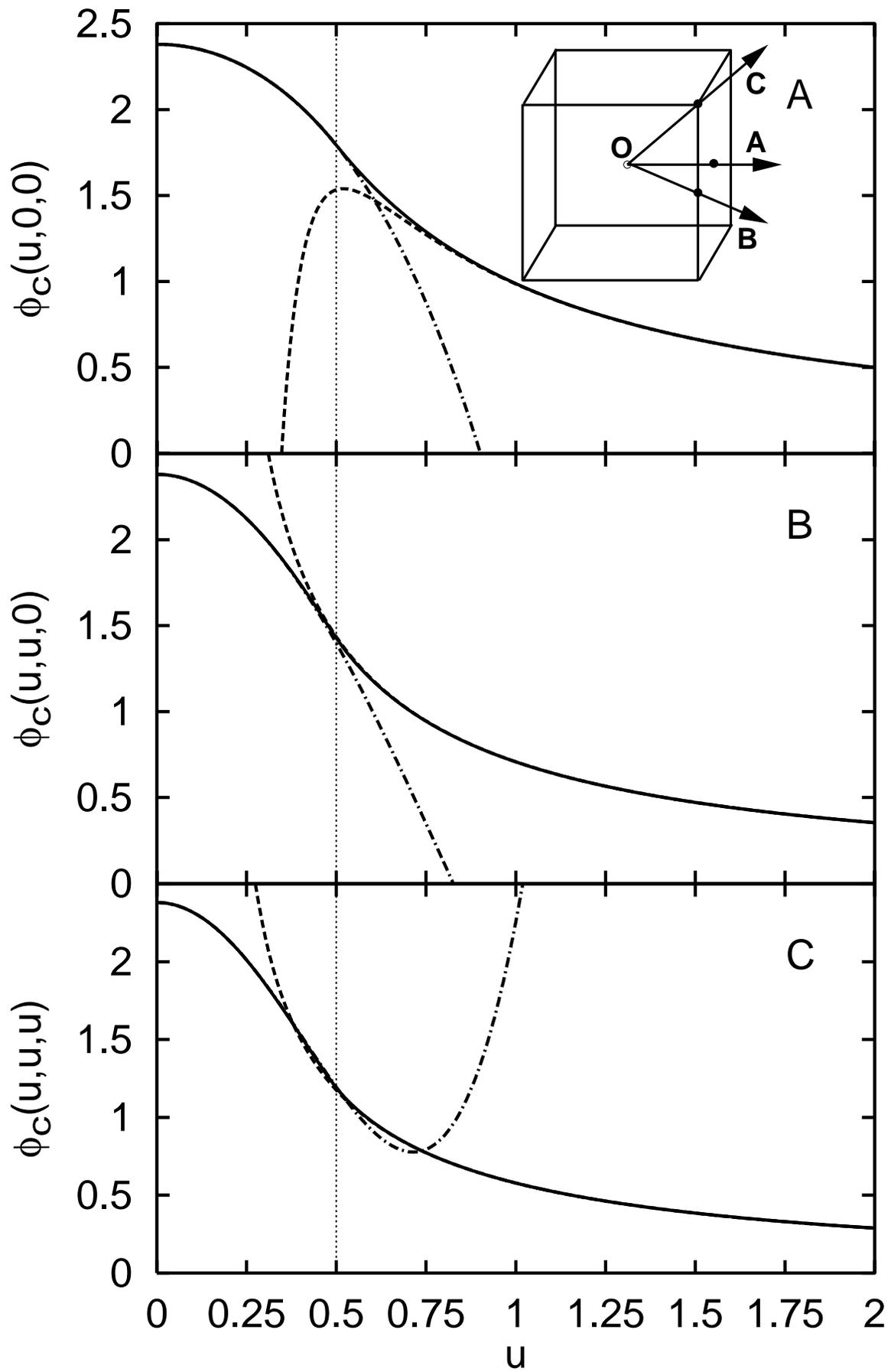
\begin{figure}[h]
\caption{
Electrostatic potential $\phi_c(u,v,w)$ of a homogeneously charged
cube. Panels A, B, and C show $\phi_c$ for directions $(u,0,0)$,
$(u,u,0)$, and $(u,u,u)$ as a function of $u$, respectively. The
insert in panel A shows a unit cube illustrating the directions A, B,
and C. The solid line is the exact result Eq.~(\protect\ref{eq:c3}).
The dashed and dot-dashed lines are the expansions
Eq.~(\protect\ref{eq:mout}) and (\protect\ref{eq:min}) up to $K_4$ for
the exterior and interior, respectively. The vertical dotted line at
$u=0.5$ indicates the boundary of the cube.}
\label{fig:axis}
\end{figure}

For the interior of the cube, we derive the multipole-expansion
coefficients of order 2, 4, and 6 from a direct Taylor expansion in
$x$ direction. The angular dependence can then be inferred by cubic
symmetry.\footnote{Some higher-order kubic harmonics are degenerate
\cite{Slattery:80}, requiring two independent expansion directions to
get the correct angular dependence.} The electrostatic potential on
the $x$ axis can be expressed as
\begin{eqnarray}
\phi_c(u,0,0) & = & \int_{-1/2-u}^{1/2-u} dx \; f(x)~,
\end{eqnarray}
where
\begin{eqnarray}
f(x) & = & \int_{-1/2}^{1/2} dy \int_{-1/2}^{1/2} dz
\left( x^2 + y^2 + z^2 \right)^{-1/2}\nonumber\\
& = & 2 \ln \frac{(4x^2+2)^{1/2}+1}{(4x^2+2)^{1/2}-1} - 2\;x\;
\arctan\frac{4x(4x^2+2)^{1/2}}{16x^4+8x^2-1}~.
\end{eqnarray}
Taylor expansion of $\phi_c(u,0,0)$ around $u=0$ yields the expansion
\begin{eqnarray}
\phi_c(\mbox{\bf r}) & = & 3 \ln \left( 3^{1/2} + 2 \right) - \frac{\pi}{2}
-\frac{2\pi}{3} \; r^2 - \frac{40}{243^{1/2}} K_4(\mbox{\bf r})
- \frac{308}{19683^{1/2}} K_6(\mbox{\bf r}) + \cdots~. \label{eq:min}
\end{eqnarray}
Figure~\ref{fig:axis} shows the electrostatic potential $\phi_c$ along
the directions $(u,0,0)$, $(u,u,0)$, and $(u,u,u)$ calculated from the
exact result Eq.~(\ref{eq:c3}) and the expansions Eq.~(\ref{eq:mout})
and (\ref{eq:min}), both including terms up to $K_4$. The expansions
show the largest disagreement near the surface of the cube ($u=1/2$)
where they start to diverge. Otherwise, they closely reproduce the
exact potential [Eq.~(\ref{eq:mout}) for $r\rightarrow\infty$ and
Eq.~(\ref{eq:min}) for $r\rightarrow 0$].

The divergent behavior reflects an inherent problem of the near- and
far-field expansions. By construction, the Laplacians of
Eqs.~(\ref{eq:mout}) and (\ref{eq:min}) are a delta function at $r=0$
and a constant $-4\pi$, respectively, independent of the order of the
expansions. The former corresponds to a unit point charge and is
correct only outside the cube; the latter corresponds to a homogeneous
charge density and is correct only inside the cube.

\section{Conclusion}\indent

Nijboer and Ruijgrok \cite{Nijboer:88} analyzed the difference between
the energy per particle in a Wigner lattice and the energy of a point
charge in the field of the other charges. These authors studied an
infinite replication of neutral cubes consisting of a unit point
charge at the center and a compensating background. A reduction of the
electrostatic potential $\phi_c$ of a homogeneously charged cube to a
one-dimensional integral resulted in
\begin{eqnarray}
\phi_c(u,v,w) & = & \frac{\pi}{8} \int_{0}^\infty dt \; t^{-2} \;
\frac{\partial}{\partial t} \left[ h(u,t) h(v,t) h(w,t) \right]~,
\label{eq:NR}
\end{eqnarray}
where
\begin{eqnarray}
h(x,t) & = & \mbox{erf}\left[\left(x+\frac{1}{2}\right)t\right]
- \mbox{erf}\left[\left(x-\frac{1}{2}\right)t\right]
\end{eqnarray}
and $\mbox{erf}$ is the error function.\footnote{Eq.~(2.6) of
Ref.~\cite{Nijboer:88} is missing a factor $\pi$ on the right-hand
side. Eq.~(2.8) of Ref.~\cite{Nijboer:88} has the correct pre-factor.}
The solution of the one-dimensional integral in Eq.~(\ref{eq:NR})
would give the closed form Eq.~(\ref{eq:c3}) of this work.
Eq.~(\ref{eq:c3}), numerical integration of Eq.~(\ref{eq:NR}), and
direct Monte Carlo integration of Eq.~(\ref{eq:int}) were compared for
a few hundred points and gave identical results within the error
margins of the numerical integration in Eq.~(\ref{eq:NR}) and the
statistical errors of the Monte Carlo procedure. Eq.~(\ref{eq:c3}) has
the advantage of being analytical. It can be evaluated fast and with
arbitrary precision on the computer.

The electrostatic potential $\phi_c(0,0,0)$ at the center of the cube
as listed in Eq.~(\ref{eq:phinull}) can be used to correct effectively
for finite-size effects in computer simulations of ionic systems under
periodic boundary conditions, when minimum-image electrostatics is
used \cite{Allen:87}. An example is the calculation of single-ion
chemical potentials
\cite{Sloth:90,Sorensen:91,Hummer:93,Hummer:95:c,Figueirido:95}, where
the electrostatic energy of an excess ion has to be calculated. The
system-size dependence is greatly reduced if the excess charge is
compensated with a homogeneous background. The electrostatic energy
$u$ of the excess charge $q$ at ${\bf r}=0$ is then the sum of the
interactions with the other charges $q_i$ at $\mbox{\bf r}_i$ and
with the background,
\begin{eqnarray}
u & = & q \sum_{i=1}^{N} q_i / r_i + q^2 \phi_c(0,0,0)/L~,
\end{eqnarray}
where a cubical box of length $L$ is used.

Another application is the calculation of electrostatic potentials
when charges are given on a grid, for instance, when ionic density
distributions are known \cite{Klement:91}. Usually, the grid charges
are assumed to be point charges. In an improved description, the
charges are smeared out over the grid cells. The electrostatic
potentials can then be calculated using Eq.~(\ref{eq:c3}) or the
multipole expansions Eq.~(\ref{eq:mout}) and (\ref{eq:min}). This
eliminates the singularities in the electrostatic potential and gives
a more accurate description near local charge concentrations.

\begin{thebibliography}{10}

\bibitem{Nijboer:88}
B.~R.~A. Nijboer and T.~W. Ruijgrok,
\newblock On the energy per particle in three- and two-dimensional {Wigner}
  lattices,
\newblock { J. Stat. Phys.}, 53 (1988) 361--382.

\bibitem{Sloth:90}
P.~Sloth and T.~S. {S{\o}rensen},
\newblock {Monte} {Carlo} calculations of chemical potentials in ionic fluids
  by application of {Widom's} formula: Correction for finite size effects,
\newblock { Chem. Phys. Lett.}, 173 (1990) 51--56.

\bibitem{Sorensen:91}
T.~S. {S{\o}rensen},
\newblock Error in the {Debye-H\"{u}ckel} approximation for dilute primitive
  model electrolytes with {Bjerrum} parameters of 2 and ca. 6.8 investigated
  by {Monte} {Carlo} methods,
\newblock { J. Chem. Soc. Faraday Trans.}, 87 (1991) 479--492.

\bibitem{Hummer:93}
G.~Hummer and D.~M. Soumpasis,
\newblock Correlations and free energies in restricted primitive model
  descriptions of electrolytes,
\newblock { J. Chem. Phys.}, 98 (1993) 581--591.

\bibitem{Hummer:95:c}
G.~Hummer, L.~R. Pratt, and A.~E. {Garc\'{\i}a},
\newblock On the free energy of ionic hydration,
\newblock { J. Phys. Chem.} (in press, 1995, chem-ph/9505005).

\bibitem{Figueirido:95}
F.~Figueirido, G.~S. {Del Buono}, and R.~M. Levy,
\newblock On finite-size effects in computer simulations using the {Ewald}
  potential,
\newblock { J. Chem. Phys.} (submitted, 1995, chem-ph/9505001).

\bibitem{Durand:64:cube}
E.~Durand,
\newblock { {\'{E}lectrostatique}}, Vol.~1, Chap.~IV.1,
\newblock Masson, Paris, 1964.

\bibitem{vdLage:47}
F.~C. {von der Lage} and H.~A. Bethe,
\newblock A method for obtaining electronic eigenfunctions and eigenvalues in
  solids with application to sodium,
\newblock { Phys. Rev.}, 71 (1947) 612--622.

\bibitem{Slattery:80}
W.~L. Slattery, G.~D. Doolen, and H.~E. DeWitt,
\newblock Improved equation of state for the classical one-component plasma,
\newblock { Phys. Rev. A}, 21 (1980) 2087--2095.

\bibitem{Adams:87}
D.~J. Adams and G.~S. Dubey,
\newblock Taming the {Ewald} sum in the computer simulation of charged
  systems,
\newblock { J. Comput. Phys.}, 72 (1987) 156--176.

\bibitem{Allen:87}
M.~P. Allen and D.~J. Tildesley,
\newblock {Computer Simulation of Liquids},
\newblock Clarendon Press, Oxford, 1987, p.~28.

\bibitem{Klement:91}
R. Klement, D.~M. Soumpasis and T.~M. Jovin,
\newblock Computation of ionic distributions around charged
  structures: {Results} for right-handed and left-handed {DNA},
\newblock {Proc. Natl. Acad. Sci. USA}, 88 (1991) 4631--4635.

\end{thebibliography}
\end{document}